\documentclass{cmbe24} 
\usepackage{amsmath,amsfonts,amssymb}
\usepackage{graphicx,wrapfig}
\usepackage[hyphens]{xurl}
\usepackage{hyperref}

\title{The Impact of Aortic Valve Stenosis on Pulse Wave Morphology: An \textit{in silico} study with 16,038 virtual subjects}

\author[1]{Robert D Wilson}
\author[1]{Sara Vardanega}
\author[1]{Jiajie Chen}
\author[2]{Lucas O Müller}
\author[1*]{Rachel E Clough}
\author[1*]{Jordi Alastruey}
\affil[1]{Biomedical Engineering, King’s College London, St Thomas’ Hospital, 249 Westminster Bridge
Rd, London, SE1 7EH, United Kingdom, \texttt{Robert.d.wilson@kcl.ac.uk}}
\affil[2]{Department of Mathematics, University of Trento, Via Sommarive 14, 38123, Trento, Italy \newline * These authors contributed equally to this study}

\summary{Aortic valve stenosis (AVS) presents challenges in asymptomatic detection, resulting in delayed intervention. This study aims to understand how AVS affects pulse wave (PW) morphology. A PW database of 16,038 virtual subjects aged 50 to 75 was created, representing normal physiology and varying AVS degrees. All subjects were simulated using a closed-loop one-dimensional/zero-dimensional blood flow model of the entire cardiovascular system, incorporating a four-chamber heart model capable of simulating different levels of AVS by reducing the orifice area of the aortic valve. Even in cases below clinical significance, distinct PW morphology changes were observed, suggesting potential for early AVS detection using peripheral PWs from non-invasive at-home devices.}

\keywords{aortic valve stenosis,  pulse wave, database of virtual subjects, computational fluid dynamics}

\begin{document}

\section{Introduction}
Aortic valve stenosis (AVS) is the most prevalent form of heart valvular disease globally, expected to dramatically increase in the US alone from 55 million in 2020 to 72 million in 2030 due to an ageing population~\cite{thaden2014global}. It often progresses rapidly from asymptomatic to symptomatic AVS, associated with high morbidity and mortality~\cite{goody2020aortic}. Early recognition and intervention in asymptomatic cases are therefore key to patient outcomes~\cite{kanwar2018management}. AVS inevitably progresses from mild cases, with varying rates among patients, and symptoms do not correlate with the rate of progression ~\cite{kanwar2018management}. 
\\\\
One-dimensional (1-D) blood flow modelling is well suited for studying pulse waveforms (PWs) under controlled variations in cardiovascular properties, such as changes in aortic valve orifice area. Obtaining \textit{in vivo} data is challenging, but 1-D modelling offers similar results at lower cost, enabling a larger dataset that is more applicable for PW analysis, including machine learning. The aim of this study is to create a dataset of simulated PW signals representing subjects aged 50 to 75 under normal physiological conditions and with varying degrees of AVS, using existing methodologies and a state-of-the-art heart model to implement AVS progression. 

\begin{figure*}[!h]
    \centering
    \hspace*{-0cm}\includegraphics[width=15cm, height=25cm, keepaspectratio]{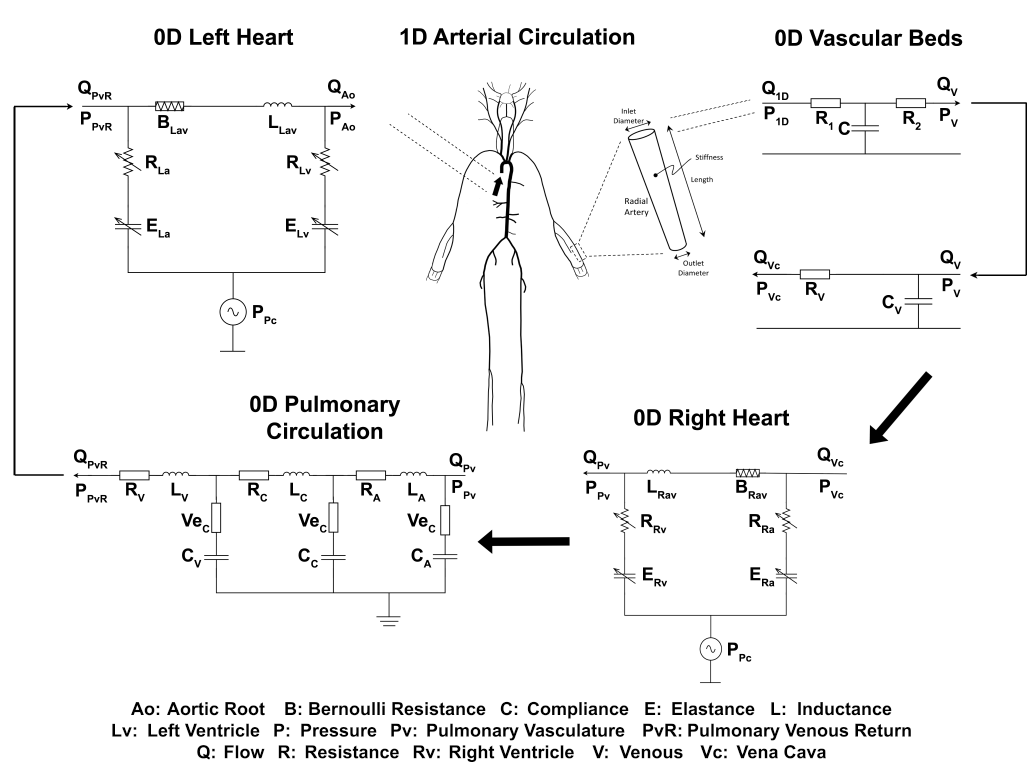}
    \caption{\textsf{The model consists of the arterial circulation of 116 larger arteries simulated using a one-dimensional (1-D) model, with the left heart chambers simulated using a zero-dimensional (0-D) replacing the original inflow waveform present at the aortic root. Lumped boundary conditions are replaced with a 0-D model description of the vascular beds at the end of each terminal 1-D model arterial segment. The heart model further including 0-D modelling of the right heart and pulmonary compartment.}}
    \label{fig:ds}
\end{figure*}

\section{Methodology}
\subsection{Generating a pulse wave dataset}
PW signals for blood pressure, photoplethysmography (PPG), blood flow velocity, and blood flow rate were simulated using one-dimensional (1-D) blood flow modelling in the larger systematic arteries of the thorax, limbs, and head, as per (Charlton et al., 2019) ~\cite{charlton2019modeling}. Coupled with a zero-dimensional (0-D) four-chamber heart model (Mynard et al., 2015) ~\cite{mynard2015one}, a closed-loop 1-D/0-D blood flow model of the entire cardiovascular system (Figure 1) was used to generate the \textit{in silico} PW dataset under normal physiological conditions. Baseline subjects aged 50 - 75 at five-year intervals were initially created. With the addition of the 0-D four-chamber heart model, adjustments were made to cardiac parameters, focusing on the age range most associated with AVS and valvular replacement (50 – 75 years). Subsequent adjustments were made to mimic haemodynamic indices from (McEniery et al., 2005) ~\cite{mceniery2005normal} and (Charlton et al., 2019) ~\cite{charlton2019modeling}, including changes in cardiac cycle duration, volume source, valve opening time, and wall elastance. 

Following the generation of baseline subjects, five cardiovascular properties were independently varied by (+-1SD) of their mean value to create the dataset under normal physiological conditions. These properties were chosen based on sensitivity analysis, resulting in 1,458 subjects under normal physiology. 

\subsubsection{Aortic valve stenosis simulation}
For each virtual subject, 10 different levels of AVS were simulated. AVS was simulated by decreasing the orifice area of the aortic valve in the 0-D left heart compartment. Under normal physiological conditions, each virtual subject had a baseline orifice area of 4.9 cm² which was decreased within the range of 3.0 cm² to 0.5 cm², in ten 0.25 cm² increments, to simulate between mild to severe AVS, according to clinical guidelines (Sha et al., 2023) ~\cite{shah2023national}. This new range of orifice areas extended the 1,458 virtual subjects under normal physiological conditions to a total of 1,458 x 11 = 16,038 subjects; 14,580 of which had some level of AVS present. Simulations were run for multiple cardiac cycles until a periodic solution was attained.

\subsection{Dataset verification}
The PW database was verified by comparing \textit{in vivo} PWs across two datasets with those simulated in healthy subjects, following the methodology described in (Charton et al., )~\cite{charlton2019modeling}. PW shapes and haemodynamic characteristic from simulated subjects were compared, at different ages, from corresponding \textit{in vivo} data from the literature ~\cite{mceniery2005normal}~\cite{fluck2014effects}.

\section{Results and Discussion}
Haemodynamic PW parameters from simulated PWs, including stroke volume, peak flow time, systolic blood pressure, pulse pressure, and pulse wave velocities, exhibited significant changes with age, mirroring \textit{in vivo} observations. Decreased orifice area led to increased time and magnitude of the aortic peak flow, along with reduced reverse aortic flow volume, resulting in more rounded flow wave in the ascending aorta with smaller peak value and reduced reverse flow (Figure 2). 

These changes in aortic PWs resulted in significant variations in peripheral PW morphology due to AVS (Figure 2) which were in agreement with \textit{in vivo} data~\cite{vennin2021novel}~\cite{flores2021estimating}~\cite{millasseau2000noninvasive} : 1) brachial pulse pressure decreased, leading to a decrease in pulse amplification; 2) peak and mean flows in the brachial and radial arteries decreased; 3) time from pulse onset to peak photoplethysmogram increased at the wrist, along with the diastolic peak/ inflection point. Strong correlations were observed between AVS severity and several PW indices calculated in peripheral locations, including the arm, which could potentially be used for early diagnosis and monitoring of AVS.

\begin{figure*}[!h]
    \centering
    \hspace*{-0.5cm}\includegraphics[width=15cm, height=25cm, keepaspectratio]{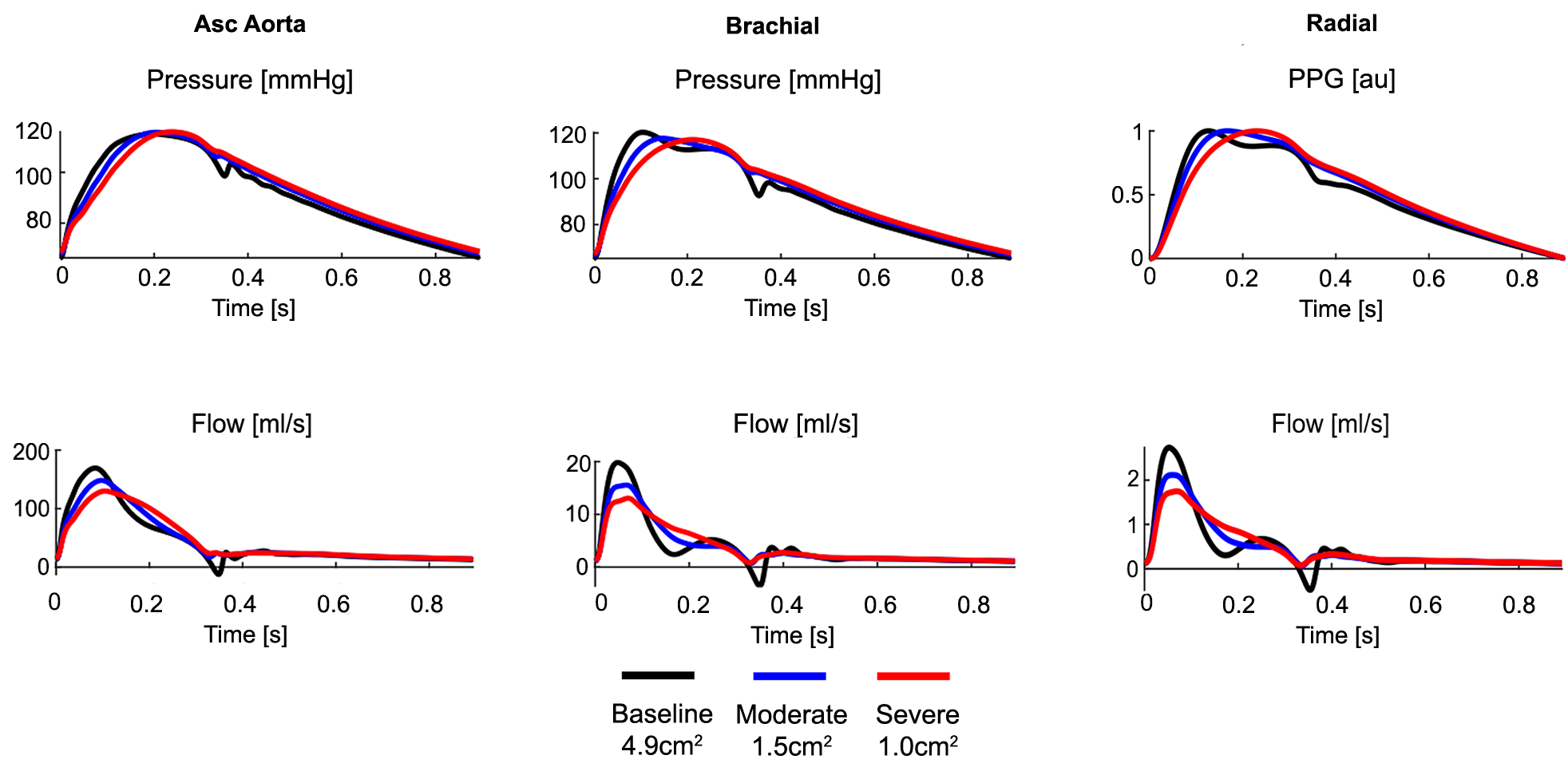}
    \caption{\textsf{Simulated pressure and flow pulse waves at the ascending aorta and brachial sites, along with the wrist
photoplethysmogram (PPG) and radial artery flow wave, for three aortic valve orifice areas for the 75-year-old baseline
subject.}}
    \label{fig:ds}
\end{figure*}

\section{Conclusions}
We have adapted and further verified an approach for simulating PWs representative of healthy adults and adults with varying levels of AVS. A 1-D computational model of the arterial system, combined with a 0-D four-chamber heart model was used to simulate flow, pressure and area PWs at common measurements sites for over 16,000 virtual subjects. Subjects of different ages were simulated through adjustments to model inputs, mimicking \textit{in vivo} vascular parameters, with the inclusion of left ventricle orifice area adjustments to simulate AVS. PWs in the database exhibit age-related haemodynamics and PW morphology representative of \textit{in vivo} data of the same age. This is the first large dataset of \textit{in silico} AVS-related PWs, which, due to its size, variance coverage, and novelty, has the potential to be used in machine learning approaches and to offer new insights into the direct non-invasive quantification of valve orifice area. 


\end{document}